\documentstyle[12pt, amsfonts]{article}
\begin{document}



\def\theequation{\thesection.\arabic{equation}}
\def\be{\begin{equation}}
\def\ee{\end{equation}}
\def\ba{\begin{eqnarray}}
\def\ea{\end{eqnarray}}
\def\lb{\label}
\def\nn{\nonumber}

\def\a{\alpha}
\def\b{\beta}
\def\g{\gamma}
\def\d{\delta}
\def\e{\varepsilon}
\def\l{\lambda}
\def\r{\rho}
\def\s{\sigma}
\def\o{\omega}
\def\O{\Omega}
\def\L{\Lambda}
\def\E{{\cal E}}
\def\Vp{{\cal V}_p}
\def\Hp{{\cal H}_p}
\def\p{\partial}
\def\bq{\overline{q}}
\def\bd{\bf d}
\def\bJ{{\bf J}}
\def\f{\frac{\pi}{k}\,}
\def\2f{\frac{2\pi}{k}\,}
\def\4f{\frac{4\pi}{k}\,}
\def\sbJ{^*\!{{\bf J}}}
\def\ux{\underline x}
\def\id{\mbox{\rm 1\hspace{-3.5pt}I}}
\newcommand{\ID}[2]{\id^{| #1 {\cal i}}_{\;\;\; {\cal h} #2 |}}

\def\C{\mathbb{C}}
\def\Z{\mathbb{Z}}
\def\F{\\mathbb{F}}
\def\1{1\!\!{\rm I}}
\def\eod{\phantom{a}\hfill \rule{2.5mm}{2.5mm}}

\def\hF{\hat{F}}
\def\hA{\hat{A}}
\def\hB{\hat{B}}
\newcommand{\dy}[1]{DYBE(${#1}$)}
\newcommand{\xR}[1]{\ ^{\small #1}\!\!\hR}
\newcommand{\xF}[1]{\ ^{\small #1}\!\!\hF}

\newcommand{\rank}{\mathop{\rm rank}\nolimits}
\newcommand{\height}{\mathop{\rm height}\nolimits}
\newcommand{\aut}{\mathop{\rm Aut}\nolimits}
\newcommand{\rx}{\mathop{\rho_{\hspace{-1pt}\scriptscriptstyle W,k}}\nolimits}
\newcommand{\rn}{\mathop{\rho_{\hspace{-1pt}\scriptscriptstyle W,n}}\nolimits}
\newcommand{\q}[1]{[#1]}
\newcommand{\ai}[1]{{a_{#1}}\, }
\newcommand{\ainv}[2]{({a^{-1})}^{| #1 {\cal i}}_{\;\;\; {\cal h} #2 |}}

\def\R{\Bbb R}

\def\Rp{\hat{R}(p)}
\newcommand{\DR}[1]{\hat{R}_{#1}(p)}
\newcommand{\DDR}[2]{\hat{R}^{#1}_{#2}(p)}
\newcommand{\DDDR}[2]{\hat{R}^{#1}_{#2}(p')}

\def\A{{A}}

\def\eup{\varepsilon^{|1  \dots n{\cal i}}}
\def\edo{\varepsilon_{{\cal h}1  \dots n|}}
\def\eupp{\varepsilon^{|1  \dots n{\cal i}}(p)}
\def\edop{\varepsilon_{{\cal h}1  \dots n|}(p)}

\def\eu2{\varepsilon^{|2  \dots n{+}1{\cal i}}}
\def\ed2{\varepsilon_{{\cal h}2  \dots n{+}1|}}
\def\eu2p{\varepsilon^{|2  \dots n{+}1{\cal i}}(p)}
\def\ed2p{\varepsilon_{{\cal h}2  \dots n{+}1|}(p)}

\def\Eup{ {\cal E}^{|1  \dots n{\cal i} } }
\def\Edo{ {\cal E}_{{\cal h}1  \dots n| } }
\def\Eupp{{\cal E}^{|1  \dots n{\cal i}}(p)}
\def\Edop{{\cal E}_{{\cal h}1  \dots n|}(p)}

\def\U2{ {\cal E}^{|2  \dots n{+}1{\cal i} } }
\def\D2{ {\cal E}_{{\cal h}2  \dots n{+}1| } }
\def\Eu2p{{\cal E}^{|2  \dots n{+}1{\cal i}}(p)}
\def\Ed2p{{\cal E}_{{\cal h}2  \dots n{+}1|}(p)}

\newcommand{\eupi}[1]{\varepsilon^{|#1 \dots n+ #1 -1{\cal i}}}
\newcommand{\edoi}[1]{\varepsilon_{{\cal h}#1 \dots n+ #1 -1|}}
\newcommand{\Eupi}[1]{{\cal E}^{|#1 \dots n+ #1 -1{\cal i}}(p)}
\newcommand{\Edoi}[1]{{\cal E}_{{\cal h}#1 \dots n+ #1 -1|}(p)}
\newcommand{\N}[2]{N^{| #1 {\cal i}}_{\;\;\; {\cal h} #2 |}}
\newcommand{\K}[2]{K^{| #1 {\cal i}}_{\;\;\;\;\;\;\;\: {\cal h} #2 |}}
\newcommand{\iK}[2]{{K^{-1}}^{| #1 {\cal i}}_{\;\;\; {\cal h} #2 |}}

\def\subbc{{\rm C}\kern-3.3pt\hbox{\vrule height4.8pt width0.4pt}\,}

\def\qd{\stackrel{.}{q}}
\def\pl{\partial}
\def\ig{\iota ({\rm tr} (gX\frac{\partial}{\partial g}) )}
\hyphenation{di-men-si-o-nal}
\begin{center}


{\Large\bf Zero modes of the $SU(2)_k\,$}\\[3 mm]
{\Large\bf Wess-Zumino-Novikov-Witten model}\\[3 mm]
{\Large\bf in Euler angles parametrization}\\[18 mm]


{\large{\bf
L. Atanasova$^{a}$,
P. Furlan$^{b,c}$
\footnote[1]{e-mail address: furlan@trieste.infn.it},
L.
Hadjiivanov$^{d,c}$
\footnote[2]{Permanent address:
Theoretical and Mathematical Physics Division,
Institute for Nuclear Research and Nuclear Energy,
BG-1784 Sofia, Bulgaria,
e-mail: lhadji@inrne.bas.bg}}},\\

$^a$Faculty of Physics, Sofia University "St. Kliment Ohridski",\\
J. Bourchier Blvd. 5, 
BG-1164 Sofia, Bulgaria\\
$^b$Dipartimento di Fisica
Teorica dell' Universit\`a degli Studi di Trieste,\\ 
Strada Costiera 11, I-34014 Trieste, Italy\\
$^c$Istituto Nazionale di Fisica Nucleare (INFN),\\
Sezione di Trieste, Trieste, Italy\\
$^d$Theoretical and Mathematical Physics Division, \\
Institute for Nuclear Research and Nuclear Energy, \\
Tsarigradsko Chaussee 72, BG-1784 Sofia, Bulgaria
\end{center}

\vspace{1cm}









\textwidth = 16truecm
\textheight = 22truecm
\hoffset = -1truecm
\voffset = -2truecm

\begin{abstract}

\noindent

We derive the Poisson brackets of the $SU(2)_k\,$
Wess-Zumino-Novi-kov-Witten chiral zero modes 
directly, using Euler angles parametrization.

\end{abstract}

\newpage

\section{Introduction}

\setcounter{equation}{0}
\renewcommand\theequation{\thesection.\arabic{equation}}

A classical mechanical model of the so called {\em zero modes} of the
chiral $SU(n)_k\,$ Wess-Zumino-Novikov-Witten (WZNW) model has been
canonically treated in \cite{FHT6}. The symplectic form for the chiral
WZNW zero modes appeared first in \cite{FG1, FG2} and later in \cite{AT}. 
These are chiral versions of what has been called the (generalized) "top" 
in \cite{F2, AF}.

The zero modes $a_C\,,\ C=L, R\,$ are
quantities of zero energy that keep track of the freedom appearing
in the standard decomposition of the $2$-dimensional
WZNW field $g(x,t)\,$ into a product of left and right moving chiral parts,
\be
\lb{ch}
g(x,t) = g_L (x^+) \, g_R^{-1} (x^-)\,,\quad x^\pm := x\pm t\,,\quad
g_C (x) = u_C (x) a_C \,,
\ee
so that
\be
\lb{Q}
g(x, 0) = u_L (x) Q u_R^{-1} (x)\,,\quad Q := a_L a_R^{-1}\,.
\ee
Classically, $g(x,t) \in G\,$ where $G\,$ is a (compact, simple) Lie group,
and (\ref{ch}) is the general solution of the equations of motion \cite{W}
\ba
&&\partial_- j_L (x,t) = 0 \quad \Leftrightarrow \quad 
\partial_+ j_R (x,t) = 0\,, \qquad 
\partial_\pm := \frac{\partial}{\partial x^\pm}\,,
\nonumber\\
&&j_L = \frac{k}{2\pi i} (\partial_+ g ) g^{-1}\,,\quad
j_R = \frac{k}{2\pi i} g^{-1} \partial_- g \,,\quad
k\,\in\,\mathbb{N}\,.
\lb{eqmotion}
\ea
It is worth mentioning here that $a_C\,,\ C=L,R\,$ can be also interpreted as intertwiners
between different initial conditions of the "classical KZ equations" for the chiral fields
(see e.g. \cite{F2}):
\be
\lb{kz}
k \,\partial_x g_L (x) =   2\pi i\, j_L (x, 0) g_L (x)\,,\quad
k \,\partial_x g_R (x) = - 2\pi i\, j_R (x, 0) g_R (x)\,.
\ee
The WZNW model being (globally) conformal invariant, it is assumed that 
$g(x+2\pi , t) = g(x,t)\,$
(the space is compact, and $t\,$ is the conformal time). The chiral fields
$g_C (x)\,$ have general group valued monodromies $M_C\,$ whereas
the monodromies $(M_p )_C\,$ of $u_C (x)\,$ are "diagonal", i.e.,
belong to the maximal torus $T_G\in G$:
\be
\lb{MC}
g_C (x + 2\pi ) = g_C (x) \, M_C\,,\quad u_C (x + 2\pi ) = u_C (x) \,
(M_p)_C\,,
\ee
so that
\be
\lb{MMp}
M_C = a_C^{-1}\, (M_p)_C \, a_C\,.
\ee
In order $g(x,t)\,$ (\ref{ch}) to be periodic in $x\,$, one should impose 
the constraint $M_L = M_R\,.$ 
The relation between $g_C (x)\,$ and $u_C (x)\,$ displayed in (\ref{ch})
(i.e., between chiral fields with general monodromy and
such with diagonal one)
is called sometimes "vertex-IRF correspondence". All this is well known;
for more details see e.g. the recent papers \cite{BFP2, GTT, FHIOPT} and references
therein.

Both $g_C (x)\,$ and $u_C (x)\,$ have quadratic Poisson brackets (PB)
involving classical $r$-matrices. The $r$-matrix appearing in the
PB of $u_C (x)\,$ \cite{BDF} (denoted in this paper as $r(p)\,$) 
is necessarily {\em dynamical} \cite{EV, HIOPT} whereas, as advocated by 
Gaw\c{e}dzki in \cite{G}, that in the PB
of $g_C (x)\,$ can be chosen to be a {\em constant} one (see \cite{BFP2,
BFP3} for a general treatment of the problem). The advantage of such a
choice is that, after quantization, it provides simple quantum group
symmetric exchange relations.

As it can be foreseen from the relation between $g_C (x)\,$ and $u_C
(x)\,$ in (\ref{ch}), the exchange relations of the quantized chiral zero
modes $a_C\,$ involve in this case both the (quantum) {\em dynamical} and
{\em constant} $R$-matrices \cite{HIOPT, HST, FHIOPT}. Taking the
quasiclassical limit, one should be able to reproduce the corresponding
classical PB for the zero modes; this consistency check has been performed
in \cite{FHT6}.

Several subtleties appear in this approach. First, the quasiclassical limit of the exchange
relations for the quantum zero modes leads to corresponding PB in which the
{\em classical dynamical} $r$-matrix $r(p)\,$ has a form which is not consistent
with the assumption that $a_C \in G\,$ for $G\,$ a (classical)
simple Lie group. Second, it is known that the classical Yang-Baxter equation (YBE)
for the {\em constant} $r$-matrix has no solutions for ${\cal G}\,$ (the Lie algebra of $G\,$)
compact (see, e.g., \cite{BFP2} for comments on this fact).

The first problem has been solved in \cite{FHT6} by assuming that
\be
\lb{deta}
\det a_C = {\cal D}_q (p) \ne 1\,
\ee
where ${\cal D}_q (p)\,$ is a specific function of
the coordinates $p\,$ of the dual ${\cal G}^*\,$ of the Lie algebra  of $G\,$ (and hence
$a_C\,$ belongs to a reductive extension of the group), and further adding certain closed
$p$-dependent $2$-form to the symplectic form (note that the latter is equivalent
to adding to the dynamical $r$-matrix $r(p)\,$ a term bilinear in the Cartan generators
which is allowed by the Etingof-Varchenko classification \cite{EV}).
The solution of the second problem can be sought along a similar way, e.g. by using
the complexification $G_{\mathbb{C}}\,$ instead of $G\,$ itself. Indeed, one is forced to use
$G_{\mathbb{C}}\,$ when applying Gaw\c{e}dzki's method of constructing a symplectic form
for $g_C (x)\,$ (and $a_C\,$) leading to PB with constant $r$-matrices \cite{G}.

The $2D$ WZNW models corresponding to {\em compact} $G\,$ at 
positive integer levels $k$
belong to the class of {\em rational} conformal field theories, and this crucial
feature is lost when going to simple noncompact or reductive groups.  
All we have to care about is, however, that
$g (x, t) \in G\,.$ Even the chiral components $g_C (x)\,$ are 
themselves "not observable", 
but in fact the condition (\ref{deta}) for 
$a_C\,$ can be compensated
by a suitable renormalization of $u_C (x)\,$ so we can safely assume that
$\det\, g_C (x)\, =\, 1\,.$
One should note that 
(a special matrix element of) the kernel $Q\,$
in (\ref{Q}) serves, after suitable 
quantization in an extended space of states,
as a generalized BRS operator whose homologies can be 
used to define a finite dimensional
"physical" subfactor. In the case of $G=SU(2)\,$ 
it has been proved that the
dimension of the latter is the right one, $k+1\,$ -- i.e., 
equals the number of
integrable representations of the affine algebra \cite{FHT2, FHT3, DVT1, DVT2}.
Thus the standard Hopf algebraic symmetry of the chiral sectors signalled by the
presence of constant $R$-matrices in the exchange relations can be treated in a way
reminiscent to the "covariant gauges" in ordinary gauge theories.

In \cite{FHT6} the $SU(n)_k\,$ WZNW zero modes' PB were derived by using an extended
phase space and a Dirac constraint procedure.  This short paper is aiming to
provide an alternative -- direct and very elementary --
derivation of the WZNW zero modes' PB in the
illustrative example of $G=SU(2)\,$ which allows a convenient parametrization
in terms of Euler angles. 

The content of the paper is the following.
In Section 2 we introduce the symplectic form for the chiral 
$SU(2)_k\,$ zero modes and compute it in terms of the Euler angle 
parameters. In Section 3 we invert the symplectic form 
to obtain the PB. The correct classical dynamical $r$-matrix
$r (p)_{12}\,$ following from the quasiclassical approximation is obtained 
in Section 4. In Section 5 the PB of the full monodromy matrix are 
derived. Finally, in Section 6 we discuss the results.

\section{Chiral symplectic form for the $SU(2)_k\,$ zero modes}

\setcounter{equation}{0}
\renewcommand\theequation{\thesection.\arabic{equation}}

The symplectic forms for the chiral WZNW zero modes
are obtained by keeping only the $a_C$-dependent parts ($C= L, R\,$)
of the symplectic forms for the chiral fields after using the decomposition
$g_C (x) = u_C (x)\, a_C\,$ (\ref{ch}). We will only consider the left
chiral part, denoting from now on $a_L =: a\,,\ (M_p)_L =: M_p\,$ (the
symplectic form for the right chiral part differs just in sign). One gets 
\cite{FG2, FHT6}
\ba
&&\O (a , M_p ) =
\O_a (a , M_p ) - \frac{k}{4\pi}\, \rho\, (a^{-1} M_p\, a)\,,
\qquad \O_a (a , M_p ) = \frac{k}{4\pi}\, \o (a , M_p )\,,
\nn\\
&&\o (a , M_p ) =
{\rm tr}\,\left( ( da a^{-1} )\, \left( 2 d M_p\, M_p^{-1} \, +\,
M_p\, ( da a^{-1}  ) M_p^{-1} \right) \right)
\lb{defO}\\
&&\rho\, (a^{-1} M_p\, a) =
{\rm tr}\, (M_+^{-1} d M_+ M_-^{-1} dM_-  )\,,\qquad M_+ M_-^{-1} =
a^{-1} M_p\, a \,.\nn
\ea
The matrices $ M_+ \,,\ M_-\,$ are upper, resp. lower triangular and
${\rm diag } (M_+ ) = {\rm diag } (M_-^{-1} )\,.$ 

The $2$-form $\O (a , M_p )\,$ is closed, due to
\be
d \rho ( M ) = \theta (M)\, := \frac{1}{3} {\rm tr}\, (M^{-1} d M )^3\,,
\qquad d \omega ( a , M_p ) = \theta (a^{-1} M_p a )\,.
\label{thMp}
\ee
Both equations in (\ref{thMp}) can be only satisfied on properly defined
open submanifolds of $G\,;$ this is well known \cite{G} and
will be made explicit in the calculations below.

In the case $a \in G = SU(2)$ we can parametrize the zero modes by Euler
angles,\footnote{A similar computation has been performed in \cite{F2,
AF} for the top. We are considering here the classical ($\hbar = 0$) but 
deformed ($k$ finite) case.}
\be
a = X B A :=
\left(\matrix{e^{i\xi}&0\cr 0&e^{- i\xi}}\right)
\left(\matrix{\cos\b &\sin\b\cr -\sin\b &\cos\b}\right)
\left(\matrix{e^{i\a}&0\cr 0&e^{- i\a}}\right)\,.
\ee
The diagonal monodromy matrix $M_p$ is given by
\be
M_p =
\left(\matrix{e^{i {\frac{\pi}{k} p}}&0\cr 0&e^{- i\frac{\pi}{k} p} }\right)
\equiv q^{- p\, h}
\lb{Mp}
\ee
where
\be
q = e^{-i \frac{\pi}{k}}
\lb{q}
\ee
is the (quasi)classical counterpart of the quantum group deformation parameter and
\be
h = \left(\matrix{1&0\cr 0&-1}\right) \equiv \s_3\,.
\lb{h}
\ee
We have the following expansion of the right invariant Lie algebra 
valued zero modes' $1$-form:
\ba
- i\, d a a^{-1} &=& \sum_{j=1}^3\, \Theta^j\, \s_j\,,
\lb{daa}\\
\Theta^1 &=& 
- \cos 2\xi\,\sin 2\b\, d\a\, +\, \sin 2\xi \, d\b
\,,\nn\\
\Theta^2 &=& 
\sin 2\xi\,\sin 2\b\, d\a\, +\, \cos 2\xi \, d\b
\,,\nn\\
\Theta^3 &=& d \xi\, +\, \cos 2\a\, d\a\,,\nn
\ea
where $\s_j\,$ are the Pauli matrices, and $\Theta^j\,$ -- the 
corresponding basic $1$-forms.

The calculation of $\o (a , M_p )$ is straightforward. 
\ba
&&2 {\rm tr}\, \left( (da a^{-1} ) d M_p  M_p^{-1}  \right) =
\frac{4 \pi}{k}\, dp \, (d\xi +  \,\cos 2\b \, d\a \,) \,,\nn\\
&&{\rm tr}\, \left( (da a^{-1} ) M_p (da a^{-1} ) M_p^{-1} \right) =\nn\\
&&= {\rm tr}\,
\left( B ( dA A^{-1} ) B^{-1} ( M_p\, ( d B B^{-1} ) M_p^{-1} - 
M_p^{-1} ( d B B^{-1} ) M_p ) \right) = \nn\\
&&= 4\, \sin \frac{2 \pi}{k}\, p\, \sin 2\b \, d \a \, d \b
\ea
or, putting the two parts together,
\be
\o (a , M_p ) = 4\, \left( \sin \frac{2 \pi}{k}\, p \,\sin 2\b \, d \a\, d \b\,
+ \frac{\pi}{k}\, dp\, ( d\xi + \,\cos 2\b \, d\a \, )\,\right)\,.
\ee
As expected, $\o (a , M_p )\,$ would be real
in the compact, $SU(2)$, case.

The full monodromy matrix is given by
\ba
 M&=&a^{-1} M_p\,a = M_+ M_-^{-1} \equiv v_0 \id + i\, 
\sum_{j=1}^3 v_j \s_j = \nn\\
    &=&\left( \begin{array}{cc}
         \cos \frac{\pi}{k}\, p + i\cos2\beta \, \sin \frac{\pi}{k}\, p &
            i\, e^{-2i\alpha}\, \sin 2\beta \, \sin \frac{\pi}{k}\, p \\[7pt]
         i\, e^{2i\alpha}\, \sin 2\beta\, \sin \frac{\pi}{k}\, p &
            \cos \frac{\pi}{k}\, p -i \cos 2\beta\, \sin \frac{\pi}{k}p\\
                        \end{array} \right)\,.
\lb{M}
\ea
Obviously, $M\,$ is unitary and unimodular for real values of the parameters; 
in particular, the real $4$-vector $(v_0 , v_1 , v_2 , v_3 )\,$ given by
\be
\lb{v}
( \cos \frac{\pi}{k}\, p ,\
\cos 2\a \cos 2\beta \, \sin \frac{\pi}{k}\, p , \
\sin 2\a \cos 2\beta \, \sin \frac{\pi}{k}\, p , \
\sin 2\beta \, \sin \frac{\pi}{k}\, p )
\ee
lies on ${\mathbb{S}}^3\,.$

To compute $\rho ( a^{-1} M_p a ) = \rho (M_+ M_-^{-1} )\,,$
let us parametrize the triangular matrices $M_\pm\,$ as
\be
\lb{Mpm}
M_+ = \left( \matrix{z&x\cr 0&z^{-1}} \right)\,,\quad
M_-^{-1} = \left( \matrix{z&0\cr y&z^{-1}} \right)\,,
\ee
where
\ba
&&xz^{-1} = i\, e^{-2i\a } \sin \f p \sin 2\b\,,\nn\\
&&yz^{-1} = i\, e^{2i\a } \sin \f p \sin 2\b\,,\nn\\
&&z^{-2} = e^{-i\f p} \cos^2 \b + e^{i\f p} \sin^2 \b = \cos \f p - i \sin \f p \cos
2\b =: e^{- i\f \d (p , \b )} ,\nn\\
&& \frac{y}{x} = e^{4i\a}\,,\quad xyz^{-2} = - \sin \2f p\, \sin^2 2\b\,.
\lb{xyz}
\ea
Then we can express
$\rho (M_+ M_-^{-1} )\,$ as
\ba
&&\rho (M_+ M_-^{-1} ) = z^{-2} \left(\, dy dx + z^{-1} d z ( y dx - x d y )
\,\right) =\nn\\
&&= \frac{1}{2} d (x y z^{-2} ) d \log \frac{y}{x}
-\, z^{-2} \left(
d (x z^{-1}) d \frac{y z^{-1}}{z^{-2}} +
d \frac{x z^{-1}}{z^{-2}} d (y z^{-1}) \right) .\
\lb{rhoxyz}
\ea
This gives
\be
\rho ( a^{-1} M_p a ) = 4 i (\sin \f p\, \sin 2\b )^2
d\a\,  d \log \frac{\cos \f p - i\sin \f p \cos 2\b }{\sin \f p \sin 2\b } \,.
\lb{rho}
\ee
Note that the $2$-form $\rho\,$ is not real.
It is easy to check, however, that the {\em external derivatives} of both
$\o\,$ and $\rho$ are equal to the {\em real} canonical
(Wess-Zumino) $3$-form $\theta\,$ which coincides,
in this case, with the volume form on $SU(2)\,$ \cite{W}.
The equality of $d\o\,$ and $d\rho\,$ ensures that $\O (a, M_p )\,$ is 
closed:
\be
\lb{theta}
d \o = d \rho = - \frac{16\pi}{k} \sin \2f p \,\sin 2\b\, d\a\, d\b\, d p \equiv \theta
\quad\Rightarrow \quad\, d \O (a , M_p ) = 0\,.
\ee

The full symplectic
form becomes very simple in this (complex) parametri-zation:
\be
\O (a, M_p ) = d ( p\, d \xi +  \d (p, \b)\, d \a ) \,.
\ee
Hence, $(\xi , p )\,$ and $( \a , \d (p, \b ) )\,,$ where
\be
\lb{delta}
\d (p , \b ) = i \,\frac{k}{\pi}\, \log \, ( \cos \f p - i \sin\f p \cos 2 
\b\, )\,,
\ee
form canonical pairs. This fact provides an independent explanation of
the chosen normalization of the matrix elements of $M_p\,$ in (\ref{Mp}) and
of $\d (p, \b )\,$ in (\ref{xyz}) (the original motivation being the quasiclassical
correspondence \cite{FHT6}).

\section{Poisson brackets}

Inverting the symplectic form, one obtains the Poisson brackets.
To get the elementary ones for the basic phase space variables,
$(q^i ) := (\xi , p, \a , \b )\,,$ we present $\O (a, M_p )\,$ as
\be
\lb{Oo}
\O (a, M_p ) = \frac{1}{2}\, \omega_{ij}\, d q^i\, d q^j\quad\Rightarrow\quad
\{ q^i , q^j \} = \o^{ij}\,,\quad \o^{ij}\, \o_{j\ell } = \d^i_\ell\,.
\ee
The result for the nonzero PB is
\ba
&&\{ \xi , p \} = 1\,,\nn\\
\lb{PB1}
&&\{ \a , \b \} = \f \, \frac{i \cos 2\b \sin \f p - \cos \f p}{2 \sin 2\b \sin \f p}\,,\\
&&\{ \xi , \b \} =\f \, \frac{\cos 2\b \cos \f p - 
i \sin \f p}{2 \sin 2\b \sin \f p}\,.\nn
\ea
Now using the Leibnitz rule we can obtain the PB for the matrix elements of
\be
\lb{amatr}
a = \left(\matrix{a^1_1 & a^1_2\cr a^2_1 & a^2_2 } \right) =
\left(\matrix{
e^{i(\xi + \a )} \cos \b &
e^{i(\xi - \a )} \sin \b \cr
- e^{i(\a -\xi )} \sin \b &
e^{- i(\xi + \a )} \cos \b } \right)\,,
\ee
the six independent ones of which are
\ba
\{a^1_1,a^1_2\}&=&\frac{\pi}{k}\,e^{2i\xi}\sin\beta\cos\beta 
=\frac{\pi}{k}\ a^1_2a^1_1\,, \nn\\  
\{a^2_1,a^2_2\}&=&-\frac{\pi}{k}\,e^{-2i\xi}\sin\beta\cos\beta=\frac{\pi}{k}\ 
a^2_2a^2_1\,, \nn\\
\lb{PB2}
\{a^1_1,a^2_1\}&=&i\frac{\pi}{k}\, e^{i\alpha}
\cot \frac{\pi}{k}p \,\sin\beta\cos\beta
 = -i\frac{\pi}{k}\cot \frac{\pi}{k}p\ a^2_1a^1_1\,,\\
  \{a^1_2,a^2_2\}&=&- i \frac{\pi}{k}\,    e^{-i\alpha} 
\cot \frac{\pi}{k}p\,\sin\beta\cos\beta
= -i\frac{\pi}{k}\cot \frac{\pi}{k}p\ a^2_2a^1_2\,, \nn\\
  \{a^1_2,a^2_1\}&=&- \frac{\pi}{k}\, \cos^2\beta\left(1+i\cot \frac{\pi}{k}p \right)=
   -\frac{\pi}{k}\, a^1_1a^2_2-i\frac{\pi}{k}\, \cot \frac{\pi}{k}p\ 
a^2_2a^1_1\,, \nn\\
  \{a^1_1,a^2_2\}&=&- \frac{\pi}{k}\, \sin^2\beta \left( 1-i\cot \frac{\pi}{k}p \right)
   =\frac{\pi}{k}\, a^1_2a^2_1-i\frac{\pi}{k}\, \cot \frac{\pi}{k}p\  
a^2_1a^1_2\,.\nn
\lb{PBaa}
\ea
Using the standard compact tensor product notations
$a_1 = a\otimes\id\,,\ a_2 = \id\otimes a\,$ etc., one can conveniently present
the above PB in the form
\be
\{ a_1 , a_2 \} = r^{\#} (p)_{12}\, a_1 a_2 - a_1 a_2 \f\, r_{12}
\lb{PBaa1}
\ee
where
\be
r_{12} = \left(\matrix{0&0&0&0\cr 0&0&1&0\cr 0& -1&0&0\cr 0&0&0&0}\right)\,,\quad
r^{\#} (p)_{12} = - i \f \cot \f p\ r_{12}
\lb{rmatrices}
\ee
are the {\em constant} and the {\em dynamical} skewsymmetric $r$-matrices, 
respectively.
Note that the imaginary unit in the expression for $r^{\#} (p)\,$ makes
the latter expressible, with real coefficients, in terms of the {\em compact}
(${\cal G} = su(2)\,$) generators,
\be
r^{\#} (p) = \2f \cot \f p\ (\tau_1\otimes\tau_2 - \tau_2\otimes\tau_1 )\,,\quad
2i\,\tau_j = \s_j\,,\quad [\tau_i , \tau_j ] = \varepsilon_{ij\ell}\, \tau_\ell
\lb{rp1}
\ee
($i,j,\ell =1,2,3\,$), i.e., $r^{\#} (p)\in {\cal G}\otimes {\cal G}\,,$
whereas the constant $r$-matrix
belongs instead to ${\cal G}_{\mathbb{C}}\otimes {\cal G}_{\mathbb{C}} =
s\ell (2)\otimes s\ell (2)\,,$
\be
r = e\otimes f - f\otimes e\,,\quad e = i\tau_1 -\tau_2\,,\quad f = i\tau_1 + \tau_2\,,
\quad [e, f ] = h\,.
\lb{r1}
\ee
The Jacobi identity of the PB (\ref{PBaa1}) is guaranteed due to the modified classical
YBE satisfied
by $r_{12}\,,$ resp. the modified classical dynamical YBE satisfied by
$r^{\#} (p)_{12}\,;$ as mentioned in the Introduction, the modified classical
YBE has no constant solutions for compact simple Lie algebras \cite{BFP2}.

The PB between the entries of $M_p\,$ and $a$ can be obtained immediately,
the only nontrivial relations being those between the (entries
of the) diagonal matrices $M_p$ and $X$:
\be
\{ M_{p1} , a_2 \} = \frac{2\pi}{k} \,\s_{12}\, M_{p1} a_2\,,\quad
\s := \frac{1}{2}\, \s_3\otimes \s_3\,.
\lb{sigma}
\ee
Note that $\s\,$ is just the diagonal of the polarized Casimir operator 
matrix,
\be
\lb{Cas}
C = e\otimes f + f\otimes e + \frac{1}{2}\, h\otimes h\,,
\ee
and that $r^{\#} (p)_{12}\,$ obeys the equation
\be
\lb{adMr}
(Ad (M_{p1}) - 1)\, r^{\#} (p)_{12} = 
\f\, (Ad (M_{p1}) + 1)\, (C_{12} - \s_{12} )\,.
\ee

Obviously, $\{ M_{p1} , M_{p2} \} = 0\,.$

\section{Changing the dynamical $r$-matrix}

The PB of the classical model described above should appear as a quasiclassical
limit of the corresponding quantum exchange algebra \cite{FHIOPT}.
This limit (see, e.g., \cite{FHT6}) essentially means that one only retains
the first order in the $\frac{1}{k}\,$ asymptotic expansion of the
quantum $R$-matrices and replaces the commutators
by Poisson brackets applying Dirac's quantization principle backwards.
The quantum dynamical $R$-matrix $R(p)_{12}\,$ has been obtained independently,
for $G=SU(n)\,,$ from the braiding properties of the suitable conformal blocks \cite{TK}
in \cite{HST} and as a solution of the quantum dynamical YBE -- in 
\cite{I2};
the corresponding "quantum matrix algebra of $SL(n)$-type" (the exchange algebra
appearing as a quantized version of the chiral zero modes) has been studied in
details in \cite{HIOPT}.

It has been shown in \cite{FHT6} that the correct classical dynamical
$r$-matrix for $G=SU(n)\,$ (of size $n^2 \times n^2\,$) obtained this way is given by
\be
r(p)^{j\ell}_{j'{\ell}'}=
\left\{
\begin{array}{ll}
\, i \frac{\pi}{k}\, {\cot} \frac{\pi}{k} p_{j\ell} \, (\d^j_{j'}
\d^\ell_{{\ell}'} -
\d^j_{{\ell}'} \d^\ell_{j'} )\ & {\rm for}\ j\ne\ell\\
\, 0& {\rm for}\ j=\ell
\end{array}
\right.\,.
\lb{correctrp}
\ee
Here $p_{j\ell} = p_j - p_\ell\,,\ j,\ell = 1,\dots ,n\,$ are coordinates in the
dual space ${\cal G}_{\mathbb{C}}^*\,$ of the Lie algebra ${\cal G}_{\mathbb{C}}\,$
so that nonnegative integer values of the $(n-1)\,$ independent of them, $p_{j\, j+1}\,,$
correspond to the $s\ell (n)\,$ dominant weights. Comparing, for $n=2\,$ and $p = p_{12}\,,$
(\ref{correctrp}) with (\ref{rmatrices}), one sees that our $r^{\#} (p)_{12}\,$ 
{\em only
reproduces the non-diagonal entries} of the correct classical dynamical $r$-matrix
$r(p)_{12}\,.$

The solution of this problem has been found in \cite{FHT6}. First, one observes that
the symplectic form $\O (a , M_p )\,$ (\ref{defO}) is invariant w.r. to rescaling of
$a\ \to \ f(p)\, a\,$ with $f(p)\,$ a scalar function (this doesn't change the
monodromy). It is easy to see that the PB for the {\em rescaled} zero mode matrix
elements get additional terms amounting to adding diagonal terms to $r^{\#}
(p)_{12}\,.$ In the general $SU(n)\,$ case, to recover exactly (\ref{correctrp}), one
also needs to add a closed $p$-dependent $2$-form to $\O (a, M_p )\,;$ in the $n=2\,$ 
case such a form does not exist since the
differential algebra on $p_{j\ell}\,,\ 1\le j < \ell \le 2\,$ is of course one 
dimensional. 

In \cite{FHT6} the computations have been made using an "extended" phase space on 
which $\det a\,$ and $P := \frac{1}{n}\, (p_1 + \dots + p_n )\,$ are not subject to 
any other conditions except regularity, $\det a \ne 0\,.$ A Dirac reduction on 
the submanifold 
\be
\lb{sub}
\det a = {\cal D}_q (p)\,,\qquad P = 0
\ee
where 
\be
{\cal D}_q (p) := \prod_{i<j}^n [p_{ij} ]\,,\qquad 
[p] := \frac{q^p - q^{-p}}{q- q^{-1}}\,,
\lb{qbra}
\ee
for $q\,$ given by (\ref{q})\,, was then shown to lead to the
quasiclassically expected PB and, in particular, to the correct 
classical dynamical $r$-matrix.

Let us compute in the $n=2\,$ case the PB of the rescaled zero modes
\be
a\quad \rightarrow \quad [p]^{\frac{1}{2}}\, a\,.
\lb{rescaled}
\ee

\medskip

\noindent
{\bf Remark.~}
The operation (\ref{rescaled}) is not innocent since, for $p \in k 
{\mathbb{Z}}\,,$ the bracket $[p]\,$ vanishes. In fact, at these values  
of $p\,,$ the function $\cot \f p\,$ is not defined as well. 
Therefore, we must exclude them from the outset, thus restricting the 
possible values of the diagonal monodromy (\ref{Mp}). It is easy to see 
that the forbidden values of $p\,$ are exactly those for which $M_p\,$ 
as well as $M\,,$ the full monodromy matrix (\ref{M}), belong to 
the center of $G\,$ (or $G_{\mathbb{C}}\,$). 

\medskip

The only nontrivial PB of the scaling factor is with the $\xi\,$ variable,
\be
\lb{pxi}
\{ \xi , [p]^{\frac{1}{2}} \} =
\,\frac{\pi}{2k}\, 
[p]^{\frac{1}{2}} \,\cot \f p\,, 
\ee
and using (\ref{PB1}) and (\ref{pxi}), one obtains, for example,
\ba
&&\{ a^1_1 , a^2_2 \} = \{ 
[p]^{\frac{1}{2}} e^{i(\xi +\a)} \cos \b , 
[p]^{\frac{1}{2}} e^{- i(\xi +\a)} \cos \b \} =\nn\\
&&= 2 i \cos^2 \b\, [p]^{\frac{1}{2}} \{ \xi , [p]^{\frac{1}{2}} \} + 
[p]\, \{  e^{i(\xi +\a)} \cos \b , e^{- i(\xi +\a)} \cos \b \} =
\lb{PBc1}\\
&&= \f\, a^1_2 a^2_1 + i \f \cot \f p\, (a^1_1 a^2_2 - a^2_1 a^1_2 )\,.\nn
\ea
The full list of the modified PB for the zero modes 
can be compactly written as 
\ba
&&\{ a^j_\a , a^j_\b \} = - \frac{\pi}{k}\ a^j_\a a^j_\b \, 
\epsilon_{\a\b} \,,\qquad j, \a , \b  = 1 , 2\,,\qquad\
 (\epsilon_{\a\b} ) = \left(\matrix{0&-1\cr 1 &0}\right)\,, 
\nn
\\
&&\{a^1_\a,a^2_\b \} =
- \frac{\pi}{k}\, ( i \cot \frac{\pi}{k}p\ ( {\rm det} a \, ) \, + 
a^2_\a a^1_\b )\, \epsilon_{\a\b} =
\lb{PBcaa}
\\
&&= - \frac{\pi}{k}\, ( i 
\frac{ \cos \frac{\pi}{k} p}{ \sin \frac{\pi}{k} } \, + 
a^2_\a a^1_\b )\, \epsilon_{\a\b} 
\nn
\ea
(no summation over $\a\,$ and $\b\,$ is assumed). The PB (\ref{PBcaa}) are 
of the form 
\be
\{ a_1 , a_2 \} = r (p)_{12}\, a_1 a_2 - a_1 a_2 \f\, r_{12}
\lb{PBaa2}
\ee
with 
\be
r (p) = r^{\#} (p) + \frac{2 \pi}{k} \cot \f p\ 
( \id\otimes \tau_3 - \tau_3 \otimes \id ) 
\lb{prc}
\ee
($r^{\#} (p)\,$ is given by (\ref{rp1})). 
The matrix 
\be
r (p)_{12} = i \f \cot \f p\, \left(\matrix{0&0&0&0\cr 0&1&-1&0\cr 
0&1&-1&0\cr 
0&0&0&0}\right)\,
\lb{rpcmatr}
\ee
coincides with (\ref{correctrp}) for $n=2\,,\ p_{12} := p = - p_{21}\,.$

Obviously, Eq.(\ref{sigma}) does not change after the rescaling.

\section{Poisson brackets for the full monodromy matrix $M$}

The PB of the zero modes with the full monodromy matrix $M\,$ as well as the PB
between the matrix elements of $M\,$ should not contain the dynamical $r$-matrix.
This conclusion follows immediately from the quasiclassical limit of the
corresponding quantum exchange algebra which does not contain the quantum dynamical
$R$-matrix, see e.g. \cite{FHIOPT}. On the other hand, using $M = a^{-1} M_p a\,$ 
(\ref{M}), one should be able to reproduce this directly. Indeed,
\ba
&&\{ M_1 , a_2 \} = \{ a^{-1}_1 M_{p1} a_1 , a_2 \} = \nonumber\\
&&= -  a_1^{-1} \{ a_1 , a_2 \} a_1^{-1} M_{p1} a_1 +
a^{-1}_1 \{ M_{p1} , a_2 \} a_1 + a_1^{-1} M_{p1} \{ a_1 , a_2 \} =\nonumber\\
&&= \frac{\pi}{k}\, a_2 r_{12} M_1 - a_1^{-1} r(p)_{12} M_{p1} a_1 a_2 
+ \frac{2 \pi}{k}\, a_1^{-1} \s_{12} M_{p1} a_1 a_2 +\nonumber\\
&&+ a_1^{-1} M_{p1} r(p)_{12} a_1 a_2 - \frac{\pi}{k}\, M_1 a_2 r_{12} 
=\nonumber\\
&&= \frac{\pi}{k}\, a_2 (r_{12} M_1 - M_1 r_{12} ) +\nonumber\\
&&+ a_1^{-1} ( M_{p1} r(p)_{12} - r(p)_{12} M_{p1} 
+ \frac{2 \pi}{k}\, \s_{12} M_{p1} ) a_1 a_2\,.
\lb{M1}
\ea
Now note that $r (p)\,$ (\ref{prc}) obeys (\ref{adMr}) together 
with $r^{\#} (p)\,$ since the diagonal elements lie in the kernel
of the operator $Ad (M_{p1}) - 1\,$ (this means, in particular, that 
the present derivation is valid for both dynamical $r$-matrices); hence,
\be
M_{p1} r(p)_{12} - r(p)_{12} M_{p1} +
\frac{2 \pi}{k}\, \s_{12} M_{p1} = 
\frac{\pi}{k}\, ( M_{p1} C_{12} + C_{12} M_{p1} )\,.
\lb{adMreq}
\ee
We have, therefore, 
\ba
&&\{ M_1 , a_2 \} = 
\frac{\pi}{k}\, a_2 (r_{12} M_1 - M_1 r_{12} )
+ \frac{\pi}{k}\,
a_1^{-1} ( M_{p1} C_{12} a_1 a_2 + C_{12} M_{p1} a_1 a_2 ) =\nonumber\\
&&= \frac{\pi}{k}\, a_2 (r_{12} M_1 - M_1 r_{12} )
+\frac{\pi}{k}\,
a_1^{-1} (M_{p1} a_1 a_2 C_{12} + a_1 a_2 C_{12} a_1^{-1} M_{p1} a_1 )
=\nonumber\\
&&= 
\frac{\pi}{k}\, a_2 (r_{12} M_1 - M_1 r_{12} ) + 
\frac{\pi}{k}\, a_2 (M_1 C_{12} + C_{12} M_1 ) =\nonumber\\
&&= \frac{\pi}{k}\, a_2 (r^+_{12} M_1 - M_1 r^-_{12} )
\lb{Mgen}
\ea
(we have used the basic property of the polarized Casimir, $[ C_{12} , a_1 a_2 ] 
= 0\,$), where
\be
r^\pm_{12}\, := \, r_{12}\, \pm \, C_{12}\,,
\lb{rpm}
\ee
i.e., we get the relation of the type we expect.

Note that the first relation in (\ref{sigma}) is equivalent to
\be
\{ a_1 , M_{p2} \} = - \frac{2 \pi}{k} \s_{12} a_1 M_{p2}\,.
\lb{sigmaeq}
\ee
Hence,
\ba
&&\{ M_1 , M_{p2} \} = \{ a_1^{-1} M_{p1} a_1 , M_{p2} \} =\nonumber\\
&&= -\,  a_1^{-1} \{ a_1 , M_{p2} \} a_1^{-1} M_{p1} a_1 + a_1^{-1} M_{p1}
\{ a_1 , M_{p2} \} =\nonumber\\
&&= \frac{2 \pi}{k}\, a_1^{-1} \s_{12} a_1 M_{p2}\, a_1^{-1} M_{p1} a_1 
- \frac{2 \pi}{k} \, a_1^{-1} M_{p1} \s_{12} a_1 M_{p2} 
=\nonumber\\
&&= \frac{2 \pi}{k} \, a^{-1}_1  
( \s_{12} M_{p2} M_{p1} - M_{p1} \s_{12} M_{p2} ) a_1 = 0
\lb{PBMMp}
\ea
($M_{p1}\,,\ M_{p2}\,$ and $\s_{12}\,$ are all diagonal and hence, commute).

We can obtain, finally, the relation for the PB $\{ M_1 , M_2 \}\,$ as well:
\ba
&&\{ M_1 , M_2 \} = \{ M_1 , a_2^{-1} M_{p2} a_2 \} =\nonumber\\
&&= -\, a_2^{-1} \{ M_1 , a_2 \} a_2^{-1} M_{p2} a_2 +
a_2^{-1} M_{p2} \{ M_1 , a_2 \} =\nonumber\\
&&= M_2 a_2^{-1} \{ M_1 , a_2 \} - a_2^{-1} \{ M_1 , a_2 \} M_2 = \nonumber\\
&&= \frac{\pi}{k}\, \left( M_2 (r^+_{12} M_1 - M_1 r^-_{12} ) - 
( r^+_{12} M_1 - M_1 r^-_{12} ) M_2 \right) =\nonumber\\
&&=  - \frac{\pi}{k}\ (M_1 M_2 r_{12} + r_{12} M_1 M_2 - M_2
r_{12}^+ M_1 - M_1 r_{12}^- M_2 )\,.
\lb{PBMM}
\ea

\section{Discussion and outlook}

Although the purpose of this work is to be as explicit as possible, many
of the computations and the comments are subject to immediate
generalization (e.g., for higher compact Lie groups). This concerns, for
example, Eq.(\ref{adMr}), which can be derived, in general, by solving the
equations for the Hamiltonian vector fields (see \cite{G}). For
$G=SU(2)\,$ the solutions (\ref{rp1}) and (\ref{prc})  of Eq.(\ref{adMr})
(both satisfying, in addition, the modified classical dynamical YBE)  
correspond to different choices for $a\,$ -- in the first case $a\,$
belongs to $G\,$ itself and in the second, to a reductive extension of
$G\,.$ As shown in \cite{FHT6}, for $SU(n)\,,\ n>2\,$ one also needs to
add a closed $p$-dependent form to the chiral symplectic form in order to
get the "correct" $r (p)_{12}\,$ (i.e., the one following from the
quasiclassical correspondence, given by (\ref{prc}) for $n=2\,$).

What characterizes the WZNW PB with {\em constant} $r$-matrices is their
Poisson-Lie invariance (see e.g. \cite{AT}) which is the classical
counterpart of the quantum group invariance of the corresponding exchange
relations for the chiral fields $g_C (x)\,$ \cite{G}. The price one has to
pay for this (relatively simple -- in particular, coassociative) symmetry
is the necessity of extending the original WZNW phase space. As mentioned
in the Introduction, the main open problem is to provide a general
procedure for finding the proper "physical quotient" of this extension. 
Qualitatively different is the situation for the chiral fields $u_C (x)\,$
with diagonal monodromy whose quantized counterparts are the chiral vertex
operators \cite{FHIOPT}. Their PB involve the dynamical §r§-matrix
$r (p)_{12}\,$ in place of $r_{12}\,.$ For a third possibility -- of
defining a {\em quasi-Poisson} structure on the chiral WZNW phase space --
see e.g. the discussion in \cite{BFP3} (and references therein). It leads,
upon quantization, to a quasi-Hopf (hence, non coassociative) deformed
symmetry \cite{Dr}. In all these cases the corresponding exchange algebras
encode the monodromy properties of the solutions of the
Knizhnik-Zamolodchikov equation for the WZNW conformal blocks.

Clearly, the zero modes contain the whole information about the
generalized WZNW internal symmetry acting from the right on the chiral
fields, cf. (\ref{ch}). They form a finite dimensional dynamical system
and hence are easier to study. The present paper is an attempt to
providing some more details clarifying the structure of the classical 
model.

\section*{Acknowledgements}

The authors thank Ivan Todorov for his valuable comments and critical
remarks.

P.F. acknowledges the support of the Italian Ministry of Education,
University and Research (MIUR). L.H. thanks Istituto Nazionale di Fisica
Nucleare (INFN), Sezione di Trieste and Dipartimento di Fisica Teorica
(DFT) dell' Universit\`a di Trieste for hospitality and support. This work
is supported in part by the EC contract HPRN-CT-2002-00325 and by contract
F-828 with the Bulgarian National Council for Scientific Research.


\end{document}